\documentclass[a4paper,10pt]{article}

\usepackage[T1]{fontenc}
\usepackage[utf8]{inputenc}
\usepackage[english]{babel}

\usepackage[text={12.36cm,19.3cm},centering]{geometry}
\usepackage{csquotes,microtype,xcolor}
\usepackage{graphicx,multirow,tikz}
\usetikzlibrary{decorations.pathreplacing}
\usepackage{amsthm,amssymb,mathtools}

\usepackage{lmodern}

\usepackage{wrapfig,caption}
\usepackage[numbers]{natbib}
\usepackage[unicode,bookmarksnumbered,colorlinks=true]{hyperref}
\hypersetup{linkcolor=blue,citecolor=blue,urlcolor=blue}

\definecolor{pd}{HTML}{495057}
\definecolor{pa}{HTML}{495057}
\definecolor{ps}{HTML}{495057}

\theoremstyle{remark}
\newtheorem{remark}{Remark}

\newcommand*{\R}{\ensuremath \mathbb{R}}
\newcommand*{\dr}{\ensuremath \partial}
\newcommand*{\intd}[1]{\ensuremath \mathrm{d}#1}
\newcommand*{\segn}{\ensuremath {\{1,\dots,n\}}}

\newcommand*{\koni}{\ensuremath k_{\text{on},i}}
\newcommand*{\koffi}{\ensuremath k_{\text{off},i}}

\title{\vspace{-18mm}\textbf{Harissa: Stochastic Simulation and Inference of Gene Regulatory Networks Based on Transcriptional Bursting}}

\author{Ulysse Herbach}

\date{\normalsize Université de Lorraine, CNRS, Inria, IECL, F-54000 Nancy, France\\\texttt{ulysse.herbach@inria.fr}}

\begin{document}

\maketitle

\begin{center}
\begin{minipage}{10.25cm}
\small\noindent\textbf{Abstract.} Gene regulatory networks, as a powerful abstraction for describing complex biological interactions between genes through their expression products within a cell, are often regarded as virtually deterministic dynamical systems. However, this view is now being challenged by the fundamentally stochastic, ‘bursty’ nature of gene expression revealed at the single cell level. We present a Python package called Harissa which is dedicated to simulation and inference of such networks, based upon an underlying stochastic dynamical model driven by the transcriptional bursting phenomenon. As part of this tool, network inference can be interpreted as a calibration procedure for a mechanistic model: once calibrated, the model is able to capture the typical variability of single-cell data without requiring ad hoc external noise, unlike ordinary or even stochastic differential equations frequently used in this context. Therefore, Harissa can be used both as an inference tool, to reconstruct biologically relevant networks from time-course scRNA-seq data, and as a simulation tool, to generate quantitative gene expression profiles in a non-trivial way through gene interactions.

\vspace{3.2mm}

\noindent\textbf{Keywords:} Gene expression\hspace{.2ex}\raisebox{-.2ex}{\textperiodcentered}\hspace{.2ex}Regulatory networks\hspace{.2ex}\raisebox{-.2ex}{\textperiodcentered}\hspace{.2ex}Single-cell data
\end{minipage}\medskip
\end{center}

\section{Introduction}

Inferring graphs of interactions between genes has become a standard task for high-dimensional statistics, while mechanistic models describing gene expression at the molecular level have come into their own with the advent of single-cell data. Linking these two approaches seems crucial today, but the dialogue is far from obvious: statistical models often suffer from a lack of biological interpretability, and mechanistic models are known to be difficult to calibrate from real data.

Here, we present a Python package for both network simulation and inference from single-cell gene expression data (typically scRNA-seq), called Harissa (\enquote*{HARtree approximation for Inference along with a Stochastic Simulation Algorithm}). It was implemented in the context of a mechanistic approach to gene regulatory network inference from single-cell data~\cite{Herbach2017} and is based upon an underlying stochastic dynamical model driven by the transcriptional bursting phenomenon. In this tool paper, we introduce briefly the main concepts behind the package, and detail its usage through some application examples.

\section{Theory}

Consider a network of $n$ genes. Our starting point is the well-known ‘two-state model’ of gene expression~\cite{Shahrezaei2008a}, which corresponds to the following set of elementary chemical reactions~\cite{Herbach2017} for each gene~$i \in \segn$:
\begin{equation}
\label{eq_reactions}
\begin{array}{ccc}
G_i \xrightleftharpoons[\koffi]{\koni} {G_i}^*,\,\quad & {G_i}^* \xrightarrow[]{s_{0,i}} {G_i}^* + X_i,\,\quad & X_i \xrightarrow[]{d_{0,i}} \varnothing , \\
& X_i \xrightarrow[]{s_{1,i}} X_i + Z_i,\,\quad & Z_i \xrightarrow[]{d_{1,i}} \varnothing ,
\end{array}
\end{equation}
where $G_i$, ${G_i}^*$, $X_i$ and $Z_i$ respectively denote ‘inactive promoter’, ‘active promoter’, mRNA and protein copy numbers for gene~$i$.
These reactions describe the main two stages of gene expression, namely \emph{transcription} (rate $s_{0,i}$) and \emph{translation} (rate $s_{1,i}$), along with degradation of mRNA (rate $d_{0,i}$) and protein (rate $d_{1,i}$) molecules. Note that $[G_i]+[{G_i}^*]=1$ is a conserved quantity: the particularity of the two-state model is that transcription of gene $i$ can only occur when its promoter is active, corresponding to $[{G_i}^*]=1$.

\begin{figure}[t]
\includegraphics[width=\textwidth]{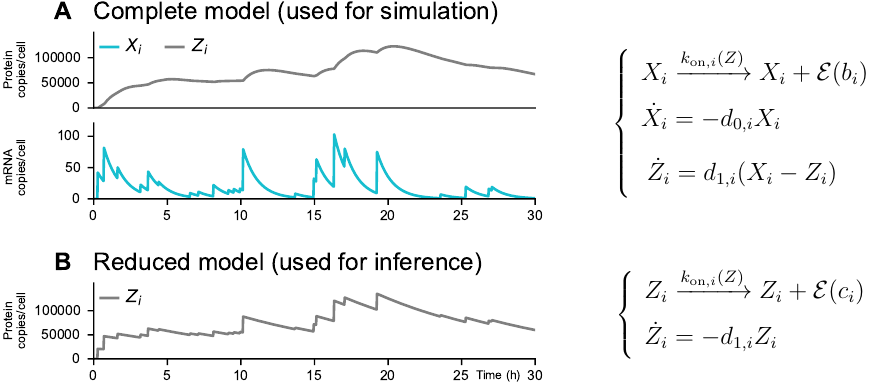}
\caption{The stochastic dynamical model underlying Harissa. (A)~Complete model with mRNA levels $X(t)$ and protein levels $Z(t)$, used to perform simulations. Transcription of gene~$i$ occurs in bursts at random times with rate $\koni(Z(t))$ and the burst size follows an exponential distribution $\mathcal{E}(b_i)$ where $b_i$ corresponds to a scaling factor. (B)~Reduced model involving only protein levels $Z(t)$. Without loss of generality, the parameter $c_i$ is set to an arbitrary value as protein levels are not measured. This model serves as a basis for deriving a pseudo-likelihood whose maximization is at the core of inference.}
\label{fig1}
\end{figure}

More specifically, experimental data consistently suggest a particular regime for this model~\cite{Tunnacliffe2020}: $\koffi \gg \koni$ and $s_{0,i} \gg d_{0,i}$ with the ratio $\koni s_{0,i}/(\koffi d_{0,i})$ remaining fixed, corresponding to short active periods during which many mRNA molecules are produced. In this regime, mRNA is transcribed by ‘bursts’ of tens to hundreds of molecules. Moreover, mRNA and protein copy numbers are not conserved quantities in the model and can be reasonably described in a continuous way using standard ‘mass action’ kinetics. This leads to a hybrid dynamical model, consisting of ordinary differential equations subject to random jumps (Fig.~\ref{fig1}) where the related quantities are denoted by $X(t) = (X_1(t),\dots,X_n(t))\in\R_+^n$ and $Z(t) = (Z_1(t),\dots,Z_n(t))\in\R_+^n$.

So far the genes are not interacting: the main point of our approach is to consider the burst frequency $\koni$ as a function of proteins levels $Z_1,\dots,Z_n$~\cite{Herbach2017}. In the current version of Harissa, this function takes the following form:
\begin{equation}
\label{eq_koni}
k_{\text{on},j}(z) = \frac{k_{0,j} + k_{1,j}\exp(\beta_j + \sum_{i=1}^n\theta_{ij}z_i)}{1 + \exp(\beta_j + \sum_{i=1}^n\theta_{ij}z_i)} \qquad\forall j\in\segn
\end{equation}
with $k_{0,j} \ll k_{1,j}$, so that $(\theta_{ij})_{1\le i,j \le n}$ can be easily interpreted as the \emph{network interaction matrix} while $\beta_j$ encodes basal activity of gene $j$ (see also Table~\ref{tab1}).
The rate $\koffi$ (inverse of mean burst duration) is kept constant.

Mathematically speaking, the complete set of variables $(X(t),Z(t))_{t\ge 0} \in \R_+^n \times \R_+^n$ forms a \emph{piecewise-deterministic Markov process} (PDMP) that is fully characterized by its (continuous) master equation:
\begin{equation}\label{eq_master_full}
\begin{array}{rll}
\displaystyle\frac{\dr}{\dr t}p(x,z,t) \hspace{-2.5mm}& \displaystyle = \sum_{i=1}^n \left[ d_{0,i}\frac{\dr}{\dr x_i}\{x_ip(x,z,t)\} + d_{1,i}\frac{\dr}{\dr z_i}\{(z_i-x_i)p(x,z,t)\} \right. & \hspace{-3mm}\vspace{2mm}\\
& \qquad\displaystyle + \left. \koni(z) \left(\int_0^{x_i} p(x-he_i,z,t) b_i e^{-b_i h} \intd{h} - p(x,z,t) \right) \right] , &\hspace{-3mm}
\end{array}
\end{equation}
where $p(x,z,t)$ is the probability distribution of $(X(t),Z(t))$. The master equation describes the time evolution of this probability and is related to the trajectory dynamics (see Fig.~\ref{fig1}A). Notably, each $Z_i$ is rescaled so that parameters $s_{1,i}$ do not appear anymore, and $s_{0,i}$ and $\koffi$ aggregate in the bursty regime into the ‘burst size’ parameter $1/b_i = s_{0,i}/\koffi$ (Table~\ref{tab1}).

\begin{table}[b]\small
\caption{Parameters of the dynamical model underlying Harissa. Here we consider an instance \texttt{model = NetworkModel(n)} where \texttt{n} is the number of genes in the network.}\label{tab1}
\hspace{-1.5mm}\begin{tabular}{r}\\
\multirow{2}{*}{\raisebox{-3.6mm}{\begin{tikzpicture}[scale=1,line width=0.8,decoration=brace]\draw[decorate,color=pd] (0,0)--(0,0.6) node [midway,left,align=right,xshift=-0.5em] {\emph{Degradation kinetics}};\end{tikzpicture}}} \\ \\
\multirow{3}{*}{\raisebox{-7.76mm}{\begin{tikzpicture}[scale=1,line width=0.8,decoration=brace]\draw[decorate,color=pa] (0,0)--(0,0.98) node [midway,left,align=right,xshift=-0.5em] {\emph{Bursting kinetics}};\end{tikzpicture}}} \\ \\ \\
\multirow{2}{*}{\raisebox{-4mm}{\begin{tikzpicture}[scale=1,line width=0.8,decoration=brace]\draw[decorate,color=ps] (0,0)--(0,0.6) node [midway,left,align=right,xshift=-0.5em] {\emph{Network parameters}};\end{tikzpicture}}} \\ \\
\end{tabular}\hspace{1mm}
\begin{tabular}{@{\hspace{.2ex}}l@{\hspace{1ex}}c@{\hspace{.5ex}}c@{\hspace{1ex}}l@{\hspace{.2ex}}}
\hline
Package variable & Notation & \hspace{0.5em} & Interpretation (gene $i$) \\
\hline
\texttt{model.d[0][i]} & $d_{0,i}$ && mRNA degradation rate \\
\texttt{model.d[1][i]} & $d_{1,i}$ && protein degradation rate \\
%\hline
\texttt{model.a[0][i]} & $k_{0,i}$ && minimal burst frequency \\
\texttt{model.a[1][i]} & $k_{1,i}$ && maximal burst frequency \\
\texttt{model.a[2][i]} & $b_i$ && inverse of mean burst size \\
%\hline
\texttt{model.basal[i]} & $\beta_i$ && basal activity \\
\texttt{model.inter[i,j]} & $\theta_{ij}$ && interaction $i \to j$\\
\hline
\end{tabular}
\end{table}

This ‘complete’ model is simulated in Harissa using an efficient acceptance-rejection method, similarly to the explicit construction given in~\cite{Benaim2015}. A great advantage of this method is that it is guaranteed to be exact without requiring any numerical integration, contrary to the basic algorithm~\cite{Malrieu2015}.

We also consider a ‘protein-only’ model (Fig.~\ref{fig1}B), which can be seen as a first-order approximation (Appendix~\ref{sec_reduced_model}): this reduced model turns out to provide a tractable inference procedure based on analytical results (Appendix~\ref{inference_algorithm}).

\section{Usage}

The Harissa package has two main functionalities: network inference interpreted as calibration of the stochastic dynamical model defined by~\eqref{eq_master_full}, and data simulation from the same model (e.g., scRNA-seq counts~\cite{Semrau2017} but also RT-qPCR levels~\cite{Richard2016,Stumpf2017}).
Besides, the package also allows for basic network visualization (directed graphs with positive or negative edge weights) as well as data binarization (using gene-specific thresholds derived from the data-calibrated dynamical model).

The first step is to create an instance of the model:
\begin{center}
\texttt{model = NetworkModel(n)}
\end{center}
where \texttt{n} is the number of genes. After setting the parameter values (see Table~\ref{tab1}), the model can be simulated using \texttt{model.simulate(time)}, where \texttt{time} is either a single time or a list of time points.

Importantly, \texttt{simulate()} does not depend on time discretization and always returns exact stochastic simulations: the resulting continuous-time trajectories are simply extracted at user-specified time points.
It is also possible to consider a stimulus, represented by an additional protein $Z_0$ that receives no feedback and verifies $Z_0(t)=0$ for $t \le 0$ and $Z_0(t)=1$ for $t > 0$. In order to reach a pre-stimulus steady state before perturbation, an optional \texttt{burnin} parameter sets the time during which the model is simulated with $Z_0(t)=0$.

\begin{wrapfigure}{r}{0.286\textwidth}\vspace{3.5mm}
\centering\includegraphics[width=0.25\textwidth]{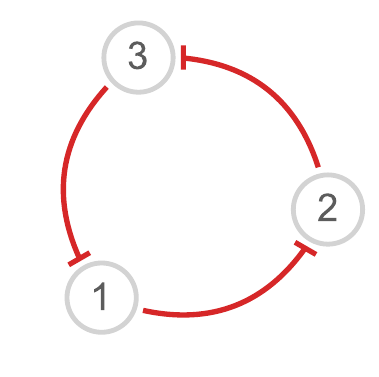}\vspace{-1.5mm}
\caption{\hspace{-0.8ex}The repressilator network used in Fig.~\ref{fig3}. This plot was made with \texttt{plot\_network} from the \texttt{harissa.utils} module.}
\label{fig2}\vspace{4mm}
\end{wrapfigure}

\paragraph{Example: Repressilator Network.}\hspace{-2mm}As an example, we consider a ‘repressilator’ network made of 3 genes forming a directed cycle of negative interactions (Fig.~\ref{fig2}).

Some critical parameters of the dynamical model are degradation rates $d_{0,i}$ and $d_{1,i}$, which characterize the ‘responsiveness’ of mRNA and protein levels. Here we set $d_{0,i}/d_{1,i} = 10$, meaning that proteins are ten times more stable than mRNAs. This is biologically realistic, but note that this ratio is known to span a very wide range~\cite{Schwanhausser2011} so there is no single choice.

\begin{figure}[t]
\includegraphics[width=\textwidth]{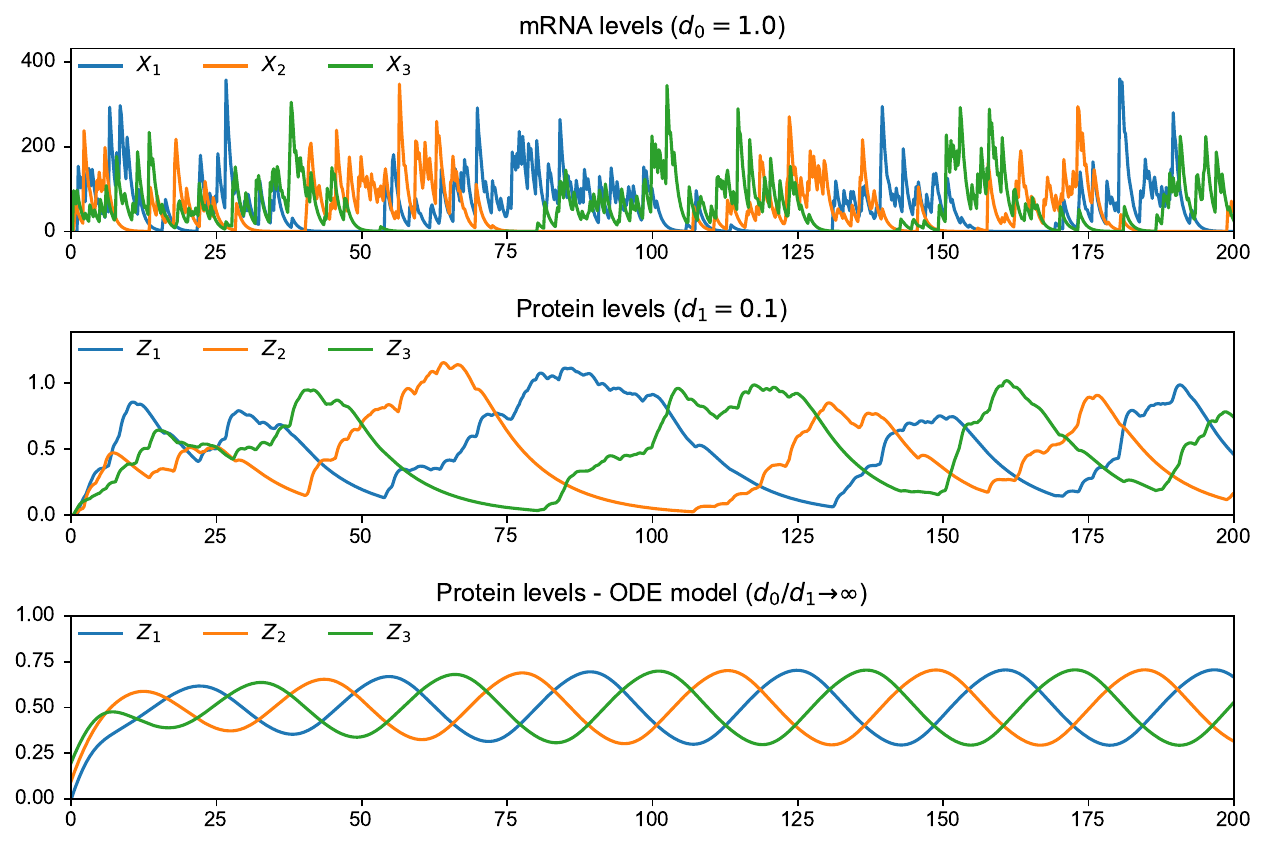}\vspace{-1mm}
\caption{Simulation of a repressilator network using Harissa. The first two plots show an example of single-cell trajectory of mRNA levels $X(t)$ and protein levels $Z(t)$ from the full stochastic model~\eqref{eq_master_full}. The bottom plot shows a trajectory of $Z(t)$ from the related deterministic model corresponding to the slow-fast limit (very stable proteins compared to mRNA, i.e. $d_0/d_1 \to \infty$). Importantly, the periodic pattern is emerging and robust well before the limit is reached (here $d_0/d_1 = 10$). The burst parameters (Table~\ref{tab1}) are $k_{0,i}=0$, $k_{1,i}=2$ and $b_i=0.02$ for all $i\in\{1,2,3\}$, which are current default values. The network parameters are given in~Appendix~\ref{sec_repressilator_network}.}
\label{fig3}
\end{figure}

An example of simulated single-cell trajectory for this network is shown in Fig.~\ref{fig3}. It is worth noticing that despite the strong level of stochasticity, a robust periodic pattern is already emerging.

More stable proteins—with respect to mRNA—will lead to less ‘intrinsic noise’ in the system. We increase mRNA degradation rates and burst frequencies instead, which is equivalent to a zoom-out regarding the time scale. Since mRNA and protein scales are normalized, the overall levels do not depend on degradation rates (but the \emph{dynamics} does).

The stochastic model converges as $d_0/d_1 \to \infty$ to a slow-fast limit~\cite{Faggionato2010} which turns out to be a nonlinear ODE system involving only proteins (bottom plot of Fig.~\ref{fig3}). It appears here that due to the deterministic dynamics, the initial protein levels need to be perturbed so as not to stay in a trivial manifold.
This limit model can be simulated with \texttt{model.simulate\_ode(time)} and is generally useful to gain insight into the system attractors (note however that it is a rough approximation of the stochastic model, see remark~\ref{rem1}).

In this slow-fast limit, which is part of the rationale behind the reduced model~(Appendix~\ref{sec_reduced_model}), mRNA levels become \emph{independent conditionally on protein levels} such that $X_i(t) \sim \mathrm{Gamma}(k_{\mathrm{on},i}(Z(t))/d_{0,i},b_i)$ for $i=1,2,3$. This quasi-steady-state (QSS) behavior can be understood intuitively from the top plot of Fig.~\ref{fig3}.
Regarding mRNA levels, \texttt{simulate\_ode} only returns the mean of the QSS distribution conditionally on protein levels (the true limit model would consist in sampling from this distribution independently for every $t > 0$).

\begin{figure}[t]\small
\resizebox{\textwidth}{!}{\begin{tikzpicture}[scale=1, line width=1]
\node (data) at (0,0) {\includegraphics[width=0.988\textwidth]{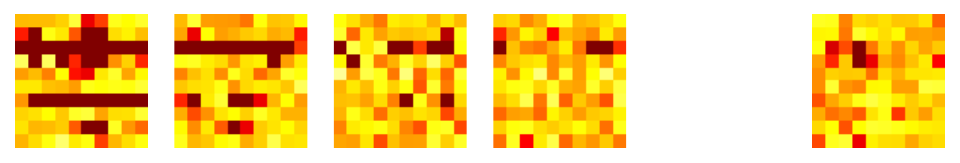}};
\node (dots) at (3.05,0) {\color{ps}\Large$\cdots$};
\draw[->,color=ps] (-5.9,1) -- (-4.25,1);
\draw[->,color=ps] (-6.06,0.825) -- (-6.06,-0.825);
\node (cells) at (-5.1,1.25) {\color{ps}\emph{cells}};
\node [rotate=90] (genes) at (-6.25,0) {\color{ps}\emph{genes}};
\draw[decorate,decoration=brace,color=ps] (-4.25,-1) -- (-5.9,-1) node[midway,below,yshift=-1mm] {$t_0$};
\draw[decorate,decoration=brace,color=ps] (-2.213,-1) -- (-3.863,-1) node[midway,below,yshift=-1mm] {$t_1$};
\draw[decorate,decoration=brace,color=ps] (-0.185,-1) -- (-1.835,-1) node[midway,below,yshift=-1mm] {$t_2$};
\draw[decorate,decoration=brace,color=ps] (1.84,-1) -- (0.19,-1) node[midway,below,yshift=-1mm] {$t_3$};
\draw[decorate,decoration=brace,color=ps] (5.895,-1) -- (4.245,-1) node[midway,below,yshift=-1mm] {$t_K$};
\end{tikzpicture}}\vspace{-1mm}
\caption{Typical structure of time-course scRNA-seq data, as required by the inference module (here each row is a gene and each column is a cell). Each group of cells collected at the same experimental time $t_k$ forms a \emph{snapshot} of the biological heterogeneity at time $t_k$. Due to the destructive nature of the measurement process, snapshots are made of different cells. This data is therefore different from so-called ‘pseudotime’ trajectories, which attempt to reorder cells according to some smoothness hypotheses.}
\label{fig4}
\end{figure}
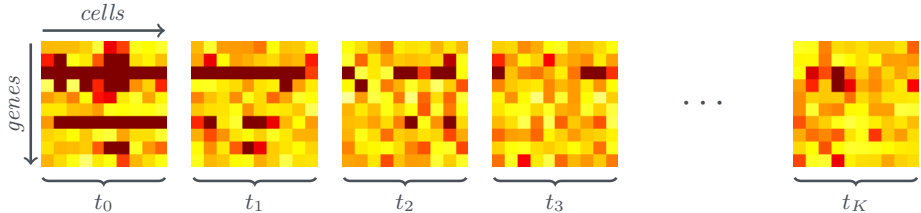

\paragraph{Network Inference.}\hspace{-2mm}Here the main function is \texttt{model.fit()}, which takes as input a time-course single-cell dataset (Fig.~\ref{fig4}). Inference can be performed by creating a new instance \texttt{model = NetworkModel()} without any size parameter, then loading a dataset \texttt{x} and using \texttt{model.fit(x)}. This will update all parameters of the model (Table~\ref{tab1}) except $d_{0,i}$ and $d_{1,i}$, which need to be provided by the user from external data—or left to their default values, see remark~\ref{rem1}—as they cannot be inferred without seriously compromising identifiability of other parameters. As an important feature, the updated \texttt{model} instance is ready for simulation, to assess reproducibility of the original data or to predict the outcome of network modifications. From a network inference viewpoint, the only important parameter is \texttt{model.inter} which is a non-symmetric signed weight matrix.

\smallskip

\begin{remark}\label{rem1}
Degradation rates $d_{0,i}$ and $d_{1,i}$ are in fact not required for the \emph{inference} part, but they are important for the \emph{simulation} part. They are currently set by default to $d_{0,i}=\ln(2)/9 \approx 0.077\;\mathrm{h}^{-1}$ and $d_{1,i}=\ln(2)/46 \approx 0.015\;\mathrm{h}^{-1}$ as median values from thousands of genes~\cite{Schwanhausser2011}. This leads in particular to $d_{0,i}/d_{1,i} \approx 5.11$.
\end{remark}

\section{Conclusions}

While the development of Harissa started a few years ago, the tool has recently become more mature and easy to use. Following early versions~\cite{Herbach2017,Herbach2018}, an alternative method for the inference part called Cardamom was developed in~\cite{Ventre2021a}, which in turn influenced the current version of Harissa. The two inference methods remain complementary at this stage and may be merged into the same package in future development. They were evaluated in a recent benchmark~\cite{Ventre2023}, giving encouraging results and showing, most importantly, that the stochastic dynamical model~\eqref{eq_master_full} can indeed reproduce real time-course scRNA-seq data through gene interactions. While the simulation part is quite well optimized and can be considered as stable, the inference part is much more challenging. Future directions include a better use of the dynamical information contained in time-course transcriptional profiles, which has great potential for performance improvement.

% Code Availability
\phantomsection
\addcontentsline{toc}{section}{Code Availability}
\paragraph{Code Availability.}\hspace{-2mm}The code of the package, a tutorial and some basic usage scripts are available at \url{https://github.com/ulysseherbach/harissa}. In addition, Harissa is indexed in the Python Package Index and can be installed via \texttt{pip}.

\appendix

\section{Appendices}

\subsection{Reduced Model}
\label{sec_reduced_model}

The inference procedure is based on analytical results which are not available for the two-stage ‘mRNA-protein’ model~\eqref{eq_master_full}. On the other hand, such results exist for a one-stage ‘protein-only’ model that is a valid approximation of the former when proteins are more stable than mRNA (i.e. $d_{0,i}/d_{1,i} \gg 1$). The resulting process $(Z(t))_{t\ge 0} \in \R_+^n$ is also a PDMP, whose master equation can be interpreted in terms of simplified trajectories (Fig.~\ref{fig1}B):
\begin{equation}\label{eq_master_reduced}
\begin{array}{rll}
\displaystyle\frac{\dr}{\dr t}p(z,t) \hspace{-2.5mm}& \displaystyle = \sum_{i=1}^n \left[ d_{1,i}\frac{\dr}{\dr z_i}\{z_ip(z,t)\} \right. & \hspace{-3mm}\vspace{2mm}\\
& \qquad\displaystyle + \left. \int_0^{z_i}\hspace{-2mm}\koni(z-he_i) p(z-he_i,t) c_i e^{-c_i h} \intd{h} - \koni(z) p(z,t) \right] . &\hspace{-3mm}
\end{array}
\end{equation}

Given $Z(t)$, mRNA levels $X(t)$ are obtained by sampling independently for every $i\in\segn$ and $t > 0$ from $X_i(t) \sim \mathrm{Gamma}(k_{\mathrm{on},i}(Z(t))/d_{0,i},b_i$), which is the \emph{quasi-steady-state} (QSS) distribution of the complete model~\cite{Herbach2017,Malrieu2015}.

\subsection{Inference Algorithm}
\label{inference_algorithm}

Now consider mRNA counts measured in $m$ cells, assumed independent, along a time-course experiment following a stimulus. Each cell $k = 1, \dots, m$ is associated with an experimental time point $t_k$. We introduce the following notation:
\begin{description}
\item $\mathbf{x}_k = (x_{ki})\in\{0,1,2,\dots\}^{n}$ : mRNA counts (cell $k$, gene $i$);
\item \vspace{-2mm}$\mathbf{z}_k = (z_{ki})\in(0,+\infty)^{n}$ : latent protein levels (cell $k$, gene $i$);
\item \vspace{-2mm}$\alpha = (\alpha_{ij}(t_k))\in\R^{n \times n}$ : effective interaction $i \to j$ at time $t_k$.
\end{description}
A stimulus is represented as gene $i=0$ and we therefore add parameters $\alpha_{0j}(t_k)$ for $j=1, \dots, n$ and $k=1, \dots, m$. We further set $z_{k0} = 0$ if $t_k \leq 0$ (before stimulus) and $z_{k0} = 1$ if $t_k > 0$ (after stimulus).
Then, writing $a_i = k_{1,i}/d_{0,i}$, the underlying statistical model of Harissa is defined by
\begin{align}
p(\mathbf{z}_k) &= \prod_{i=1}^n {z_{ki}}^{c_i \sigma_{ki} - 1} e^{-c_i z_{ki}} \frac{{c_i}^{c_i\sigma_{ki}}}{\mathrm{\Gamma}(c_i\sigma_{ki})} \;, \label{eq_stat_z}\\
p(\mathbf{x}_k | \mathbf{z}_k) &= \prod_{i=1}^n \frac{1}{x_{ki}!} \frac{\mathrm{\Gamma}(a_i z_{ki} + x_{ki})}{\mathrm{\Gamma}(a_i z_{ki})} \frac{{b_i}^{a_i z_{ki}}}{(b_i+1)^{a_i z_{ki} + x_{ki}}} \;, \label{eq_stat_xz}
\end{align}
with
\vspace{-3mm}\begin{equation}
\sigma_{ki} = {\left[1+\exp(-\{\beta_i + \alpha_{0i}(t_k) z_{k0} + \textstyle\sum_{j=1}^n \alpha_{ji}(t_k) z_{kj}\})\right]}^{-1} .
\end{equation}
Details of this derivation can be found in~\cite{Herbach2017,Ventre2021a,Herbach2019,Sarkar2021}. Roughly, \eqref{eq_stat_z} comes from a ‘Hartree’ approximation of~\eqref{eq_master_reduced}, while~\eqref{eq_stat_xz} corresponds to a Poisson distribution with random parameter sampled from the QSS distribution of $X(t)$ given $Z(t)$. Note that $p(\mathbf{z}_k)$ is in general only a pseudo-likelihood as $\sigma_{ki}$ depends on $\mathbf{z}_k$.

Since the preliminary version of Harissa~\cite{Herbach2017}, the global inference procedure has been heavily improved using important identifiability results from~\cite{Ventre2021a,Ventre2021}. The final algorithm consists of three steps:
\begin{enumerate}
\item \emph{Model calibration:} estimate $a_i$ and $b_i$ for each gene individually from~\eqref{eq_stat_xz};
\item \emph{Bursting mode inference:} estimate the frequency mode ($k_{0,i}$ or $k_{1,i}$) for each gene in each cell (can be seen as a binarization step with specific thresholds);
\item \emph{Network inference:} consider $\mathbf{z}_k$ as observed from step 2 and maximize~\eqref{eq_stat_z} with respect to $\alpha$ after adding an appropriate penalization term~\cite{Ventre2021a}. Each parameter $\theta_{ij}$ is then set to $\alpha_{ij}(t_k)$ with $t_k$ that maximizes $|\alpha_{ij}(t_k)|$.
\end{enumerate}

\subsection{Repressilator Network}
\label{sec_repressilator_network}

Considering an instance \texttt{model = NetworkModel(3)}, the repressilator network simulated in Fig.~\ref{fig3} is defined as follows:
\begin{verbatim}
model.d[0] = 1 # mRNA degradation rates
model.d[1] = 0.1 # Protein degradation rates
model.basal[1] = 5 # Basal activity of gene 1
model.basal[2] = 5 # Basal activity of gene 2
model.basal[3] = 5 # Basal activity of gene 3
model.inter[1,2] = -10 # Interaction 1 -> 2
model.inter[2,3] = -10 # Interaction 2 -> 3
model.inter[3,1] = -10 # Interaction 3 -> 1
\end{verbatim}

% Acknowledgements
\phantomsection
\addcontentsline{toc}{section}{Acknowledgements}
\paragraph{Acknowledgements.}\hspace{-2mm}The author is very grateful to Elias Ventre and Olivier Gandrillon for fruitful discussions which led to improve the Harissa package.

% References
\phantomsection
\addcontentsline{toc}{section}{References}
\renewcommand{\bibname}{References}

\end{document}